\documentclass[10pt]{iopart}
\usepackage{graphicx}
\usepackage[utf8]{inputenc} 
\usepackage[T1]{fontenc}
\usepackage{color}
\usepackage{soul} 
\usepackage[
,textwidth=15cm
,textheight=20cm
,verbose
,dvips
]{geometry}
\usepackage[numbers,comma,square,sort&compress,merge]{natbib}
\usepackage{iopams}
\usepackage[switch]{lineno} 

\newcommand{\degC}{\,$^\circ{}$C}

\graphicspath{{Figures/}}

\begin{document}
	\title[High-temperature MBE of AlN NWs]{Growth kinetics and substrate stability during high-temperature molecular beam epitaxy of AlN nanowires}
	
	\author{P.\,John$^1$, M.\,G\'{o}mez Ruiz$^1$, L.\,van Deurzen$^2$,  J.\,L\"ahnemann$^1$, A.\,Trampert$^1$, L.\,Geelhaar$^1$, O.\,Brandt$^1$, and T.\,Auzelle$^1$ }
	
	\address{$^1$ Paul-Drude-Institut f\"ur Festk\"orperelektronik, Leibniz-Institut im Forschungsverbund Berlin e.V.,	Hausvogteiplatz 5-7, 10117 Berlin, Germany}
	\address{$^2$ School of Applied and Engineering Physics, Cornell University, 14853 Ithaca New York, USA}
	
	\ead{john@pdi-berlin.de}

	
	\begin{abstract}
		We study the molecular beam epitaxy of AlN nanowires between $950$ and $1215$\degC{}, well above the usual growth temperatures, to identify optimal growth conditions. The nanowires are grown by self-assembly on TiN(111) films sputtered onto Al$_2$O$_3$. Above $1100$\degC{}, the TiN film is seen to undergo grain growth and its surface exhibits \{111\} facets where AlN nucleation preferentially occurs. Modeling of the nanowire elongation rate measured at different temperatures shows that the Al adatom diffusion length maximizes at $1150$\degC{}, which appears to be the optimum growth temperature. However, analysis of the nanowire luminescence shows a steep increase in the deep-level signal already above $1050$\degC{}, associated with O incorporation from the Al$_2$O$_3$ substrate. Comparison with AlN nanowires grown on Si, MgO and SiC substrates suggests that heavy doping of Si and O by interdiffusion from the TiN/substrate interface increases the nanowire internal quantum efficiency, presumably due to the formation of a SiN$_x$ or AlO$_x$ passivation shell. The outdiffusion of Si and O would also cause the formation of the inversion domains observed in the nanowires. It follows that for optoelectronic and piezoelectric applications, optimal AlN nanowire ensembles should be prepared at $1150$\degC{} on TiN/SiC substrates and will require an \emph{ex situ} surface passivation.
		
	\end{abstract}
	
	\vspace{2pc}
	\noindent{\it Keywords}: AlN, nanowires, molecular beam epitaxy, self-assembly, diffusion length, point defects \\
	%

	%
	\maketitle
	%
	\ioptwocol
	
	\section{Introduction}\label{sec:Intro}
	
	Ensembles of AlN nanowires (NWs) form attractive quasi-substrates for hosting optoelectronic devices operating in the deep ultraviolet (DUV) spectral range \cite{Zhao2020,liu_2021}. Compared to planar substrates, the NW geometry offers enhanced strain relaxation by deformation at the free sidewall surfaces and allows growth on both polar and nonpolar facets \cite{glas_2006a,Zhang2011}. This is useful to prevent cracks \cite{katagiri_2009}, to eliminate threading dislocations \cite{vennegues_2000,Hersee2011}, to enhance dopant incorporation \cite{Fang2015,Siladie2019}, and to grow heterostructures free of the quantum confined Stark effect \cite{coulon_2018a}. Uncoalesced NW ensembles also exhibit a higher light extraction efficiency \cite{lu_2021} and can be peeled-off for transfer onto functional substrates \cite{guan_2016}. In addition to its ultra-wide bandgap, AlN has the highest piezoelectric coefficient among binary nitride semiconductors. Piezoelectric devices based on NWs are common building blocks in sensors and nanogenerators that exhibit flexibility or require conformability \cite{Liu2022,Wang2006}. Embedding AlN NWs in these devices could be advantageous for increasing the operating temperature thanks to the superior thermal stability of AlN compared to other piezoelectric ceramics \cite{akiyama_2009}. 
	
	For these various purposes, it is desirable to fabricate AlN NW ensembles on a large scale, together with minimal structural disorder and high chemical purity. We have recently demonstrated the self-assembly of AlN NW ensembles on metallic TiN films by molecular beam epitaxy (MBE) \cite{Azadmand2020}. The NWs show intense band-edge luminescence, grow vertically, and are nearly uncoalesced due to the underlying TiN layer. 
	In contrast to other fabrication methods, no substrate pre-patterning is required and the TiN nucleation layer can be deposited on essentially any crystalline substrate \cite{Moatti2018,chang_2019}. 
	The TiN film can act as an electrode for carrier injection and has specific surface properties that can be useful for reducing the density of self-assembled NWs \cite{vanTreeck2018,Auzelle2022}.
	
	Optimum growth of AlN NWs by MBE requires elevated temperatures. For comparison, homoepitaxial layers on bulk AlN are typically grown at 1000--1100\degC{} \cite{Lee2020,Singhal2022}, and most likely even higher substrate temperatures are needed to enable the one-dimensional growth mode of catalyst-free NWs, similar to the case of GaN \cite{fernandez-garrido_2009}. Such a high temperature requirement imposes limitations on the choice of substrate. We emphasize that even if a substrate has a high melting point, solid-state reactions and interdiffusion at heterointerfaces can occur at much lower temperatures \cite{Setoyama1999,hultman_2000,Schroeder2015,yang_2018,Cancellara2021}. 
	High substrate temperatures are thus a concern for MBE growth of AlN NWs directly on GaN NW stems, a popular approach used so far for DUV devices \cite{Sarwar2015,Carnevale2013,Kent2014,Zhao2015,zhao_2017,liu_2017,Sun2018,Siladie2019,wu_2020,Vermeersch2021,cardoso_2022}, since thermal decomposition of bare GaN NWs starts already at 850\degC{} \cite{Consonni2013}. Although increased stability can be achieved by encapsulating the NWs in AlN \cite{bhunia_2020}, growth above 965\degC{} has not yet been reported \cite{wu_2020}. In comparison, our AlN NWs on TiN have been grown at nearly 1200\degC{}. However, they feature a pronounced deep-level luminescence, which has been tentatively attributed to O incorporation from the Al$_2$O$_3$ substrate. It remains to be determined which are the optimal growth conditions, in particular the substrate temperature, to balance the formation of intrinsic point defects and the incorporation of imputities.
	
	In this article, we specifically investigate the impact of substrate temperature between $950$ and $1215$\degC{} on the AlN NW growth kinetics, point defect incorporation and substrate stability. Above $1000$\degC{}, the TiN film is seen to undergo grain growth and its surface exhibits \{111\} facets where AlN nucleation preferentially occurs. Modeling of the NW elongation rate measured at different temperatures shows that the Al adatom diffusion length maximizes at $1150$\degC{}, which thus appears as the optimal growth temperature. Yet, analysis of the NW luminescence shows a steep increase in the deep-level signal already above 1050\degC{}, which is associated with O incorporation from the substrate. Comparison between AlN NWs grown on Si, MgO and SiC substrate suggests that heavy Si and O doping due to interdiffusion at the TiN/substrate interface increases the NW internal quantum efficiency, presumably by forming a SiN$_x$ or AlO$_x$ passivation shell.
	
	\section{Experimental Section} \label{sec:Exp}
	
	AlN NWs are grown by plasma-assisted MBE on 400\,nm thick TiN layers. The TiN is reactively sputtered on 2\,inch Al$_2$O$_3$(0001) substrates in a magnetron sputtering system equipped with a Ti target. The substrate is set at a temperature of 300\degC{} and biased to 50\,V. Sputtering is done for 20\,min at a pressure of 5$\times$10$^{-3}$\,mbar, using Ar and N$_2$ flows of 16\,sccm and 4\,sccm, respectively, and with a DC plasma power of 560\,W. Similar conditions are used for deposition on 1$\times$1\,cm$^2$ MgO(111),  1$\times$1\,cm$^2$ 4H-SiC(0001) and 2\,inch Si(111) substrates. The TiN films are characterized by atomic force microscopy (AFM) and by electron backscatter diffraction (EBSD). The EBSD patterns reveal a (111) orientation as confirmed by reflection high energy electron diffraction (RHEED). Exemplary RHEED patterns are shown in Fig.\,S1 of the Supporting Information (SI). After sputtering, the samples are transferred in ultrahigh vacuum to the MBE chamber where AlN NW growth takes place. The growth temperature is determined by a pyrometer working at 920\,nm and calibrated with the 1$\times$1$\leftrightarrow$7$\times$7 transition in the surface reconstruction of Si(111) occurring at 860\degC{} \cite{Hirabayashi1993,Sato1967}. The temperature-dependent emissivity of TiN is taken from Ref.\,\cite{Briggs2017}. The overall low emissivity of the TiN layer allows high growth temperatures (here up to 1215\degC{}), by limiting radiative heat dissipation from the substrate surface facing the liquid nitrogen-cooled cryo-shroud. Unless indicated differently, the TiN films are exposed to the N flux when ramping up the substrate temperature above 900\degC{} prior to AlN deposition. AlN growth proceeds for 180\,min in N-rich conditions with a N flux of 20\,nm/min and an Al flux of 2\,nm/min and a substrate rotation of 2.2\,rpm.
	The atomic fluxes are calibrated by measuring the growth rate of GaN and AlN layers in the N and Al limited regimes, respectively. The axis of the Al and N cells are 38\,$^\circ{}$ tilted with respect to the substrate normal and both cells are located on opposite sides in the azimuthal plane of the growth reactor.
	
	Once exposing the substrate to both Al and N atomic beams at high temperature, the AlN nucleation takes place after an incubation time as typically reported for self-assembled GaN NWs \cite{Consonni2011a,sobanska_2016}. This incubation time can be determined by monitoring \textit{in situ} the appearance of the first AlN nuclei by RHEED or the drop in laser reflectance, measured here at a wavelength of 650\,nm with an angle of incidence of 74\,$^\circ{}$ \cite{Corfdir2018}. The NW lengths are determined \emph{ex situ} by cross-sectional  scanning electron microscopy (SEM). The mean NW elongation rate is eventually deduced by dividing the average NW length by the growth duration (after subtraction of the incubation time). The NW growth rate is also extracted from the oscillations in the laser reflectance signal using a simple effective medium approach \cite{Corfdir2018}, as exemplified in Fig.\,S1 of the SI. To this end, the AlN fill factor $f$ in the effective layer is determined from \emph{ex situ} top-view SEM images and we assume $n_{\mathrm{eff}} = 1 + f(n_{\mathrm{AlN}} -1)$, where $n_{\mathrm{eff}}$ and $n_{\mathrm{AlN}} = 2.1$ are the refractive indexes of the effective layer and of AlN, respectively. 
	
	The presence of point defects in the AlN NWs is studied by cathodoluminescence (CL) spectroscopy performed at 10\,K in a Zeiss Ultra55 SEM with an acceleration voltage of 5\,keV and a beam current of 17\,nA. The system is equipped with a Gatan MonoCL4 for signal collection and a 1200\,lines/mm grating with photomultiplier tube for signal detection. All spectra are collected over an area of about 100\,$\mathrm{\mu m^{2}}$ and corrected with the instrumental response. Extended defects are examined by cross-section transmission electron microscopy (TEM) in a Jeol 2100F field emission microscope operated at 200\,kV and equipped with a Gatan Ultra Scan 4000 charge coupled device. The specimens were prepared by standard mechanical grinding and dimpling methods, where final thinning was achieved using a 3\,kV Ar ion beam in a Gatan precision ion polishing system at an incident angle of 3\,$^\circ{}$.

	\section{Results and discussion}\label{sec:Results}
	\subsection{TiN thermal stability}\label{subsec:TiN-results}
	
	\begin{figure*}\centering
		\includegraphics[width=.8\linewidth]{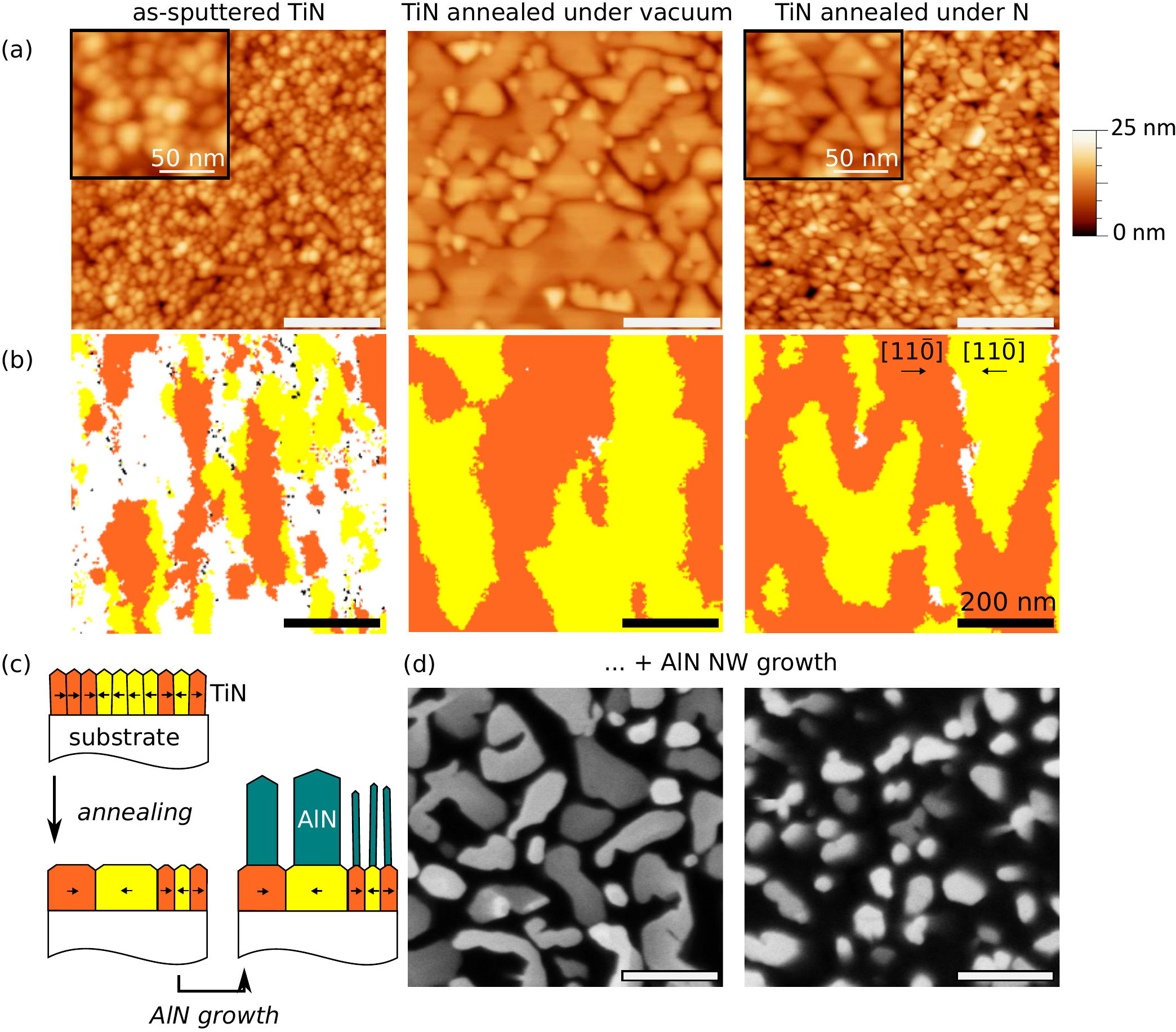}
		\caption{(a) AFM topographs and (b) EBSD maps of TiN films that are as-sputtered, and after annealing for 15\,min at 1150\degC{} in vacuum or under N exposure. The insets show magnified AFM topographs of the respective samples. Colored regions in the EBSD mapping indicate TiN grains with the [111] direction normal to the surface and an in-plane orientation as indicated in the right-most map. White areas indicate grains whose average orientation cannot be determined accurately. (c) Schematic of the sample morphology after annealing and AlN growth. (d) Top-view SEM images of AlN NWs grown on the annealed TiN films in vacuum (left) and under N flux (right). If not mentioned, the length of scale bar is 200\,nm.}\label{fig1}
	\end{figure*} 
	
	\begin{figure*}
		\includegraphics[width=.85\textwidth]{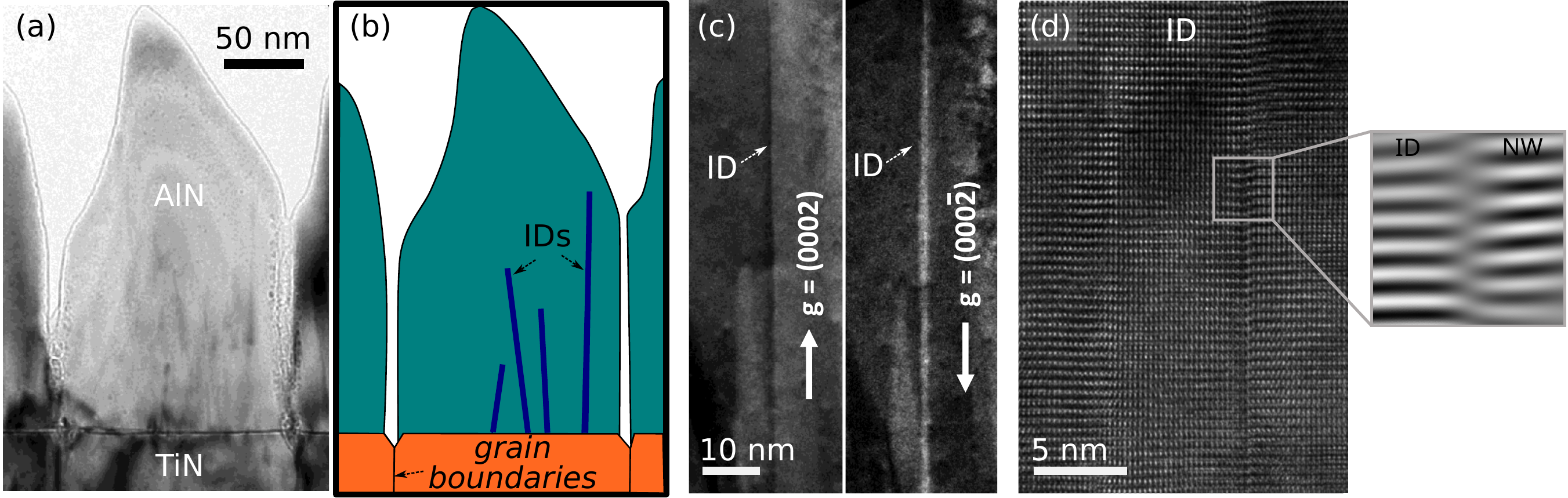}\centering
		\caption{(a) Cross-section bright-field TEM image of AlN NWs grown on a vacuum annealed TiN. (b) Schematic of the micrograph shown in (a) highlighting AlN inversion domains (IDs) and TiN grain boundaries. (c) Dark-field TEM images of an AlN ID taken under $g = (0002)$ (left) and $g = (000\bar{2})$ (right) Bragg conditions. The contrast reversal confirms the different polarities between the ID and the NW matrix. (d) High-resolution TEM image of a 7\,nm wide ID. The magnified Fourier filtered section showing (0002) lattice fringes reveals a lattice shift by half a unit cell at the ID boundary.}\label{fig2}
	\end{figure*}
	
	Although TiN exhibits a melting point of 2900\degC{}, a major reshaping of sputtered TiN layers is already observed at 1100 -- 1300\degC{} \cite{Heau1999,mader_1989,Auzelle2022}, which is likely due to the abundance of defects.
	Since the AlN NWs are grown at (1100\,$\pm$\,100)\degC{}, we examine here in detail the morphological change in the TiN layer during annealing at these temperatures, thus mimicking the conditions prior to AlN NW growth.
	
	As illustrated in Figs.\,\ref{fig1}(a), left-hand side and \ref{fig1}(c), the as-sputtered TiN surface is composed of coalesced grains exhibiting a diameter of $\approx$\,25\,nm. \emph{In situ} RHEED monitoring reveals a transmission pattern consisting of diffraction spots and weak chevrons, associated to a surface morphology characterized by TiN \{100\} facets, as discussed in our previous work \cite{Auzelle2022}. This morphology typical for sputtered TiN results from the formation of columnar grains which extend along the [111] direction, perpendicular to the substrate surface, and with some misorientation between each other \cite{Thornton1986}. As expected, TiN crystallizes in two different in-plane orientations on sapphire \cite{Smith2020,Gao2022,Moatti2018,Grundmann2011}. EBSD mapping reveals the epitaxial relationships of [111]$_\mathrm{TiN}$||[0001]$_\mathrm{Al_2}$$_\mathrm{O_3}$ and $\langle$$1\bar{1}0$$\rangle$$_\mathrm{TiN}$||[$10\bar{1}0$]$_\mathrm{Al_2}$$_\mathrm{O_3}$,where the two TiN rotational domains are shown as orange and yellow areas in Fig.\,\ref{fig1}(b). They exhibit a surface ratio of 1:1 and generally include more than 10 columnar TiN grains. White areas in the same map indicate grains where the average orientation cannot be determined accurately due to disorder on a scale smaller than the EBSD resolution.
	
	After 15\,min of annealing at 1150\degC{} in vacuum, a reshaping of the TiN film occurs, leading to an overall reduction in the texture of the film. The TiN surface is now characterized by coalesced triangular grains having a characteristic length of $\approx$\,150\,nm, a $\{111\}$ top facet, surrounded by grooves [Fig.\,\ref{fig1}(a), center]. The size increase of the triangular facets during annealing results from grain growth, a process that is driven by the motion of grain boundaries in order to reduce their strain and areal density \cite{Thompson2000}. In comparison, grain growth was observed already from $650$\degC{} for TiN reactively sputtered on stainless steel substrates \cite{Heau1999}. EBSD mapping of the annealed TiN films now resolves a clear [111] orientation for nearly all TiN grains. A similar process is observed when annealing the TiN substrate under N exposure, although smaller grains are obtained ($\approx$\,40\,nm in diameter) [Fig.\,\ref{fig1}(a), right-hand side]. It shows that the presence of N adatoms slows down grain growth, suggesting that this process is mediated by the surface. Raman spectroscopy confirmed that the TiN stoichiometry below the surface remains unchanged after annealing \cite{Auzelle2022}. We propose that the reshaping is triggered by the formation of N vacancies at the surface of the TiN grains (including grain boundaries). These N vacancies will change the surface energy \cite{lee_2012} and will ease the surface diffusion of Ti atoms \cite{vangastel_2000}, both aspects driving the layer transformation and the concomitant appearance of \{111\} facets. The vacancies can form due to desorption of two surface N atoms in the form of N$_2$ or as a result of outdiffusion of N via vacancies --- a common defect in sputtered TiN layers \cite{schaffer_1992,hultman_2000,qi_2019}. Vacancy-assisted diffusion has been found to set in at temperatures above $\approx$\,900\degC{} due to its large diffusion barrier (3.8\,eV) \cite{tsetseris_2007a}.
	
	When AlN NW growth is attempted at 1150\degC{} on the vacuum annealed TiN, immediate nucleation is observed \emph{in situ} by RHEED and laser reflectance. In contrast, the incubation time amounts to ($15\pm5$)\,min when nucleating AlN on the TiN annealed under N exposure. At this temperature, \emph{in situ} RHEED monitoring reveals a transformation to a reflection pattern consisting of vertical diffraction streaks prior to AlN nucleation, which is associated to the formation of TiN \{111\} facets \cite{Auzelle2022}. We thus conclude that AlN nucleation preferentially occurs on the \{111\} TiN facets, similar to the GaN case \cite{Auzelle2022}. At lower temperatures ($<$1100\degC{}), large \{111\} TiN facets do not form but AlN nucleation may still occur at edges between \{100\} facets. We note that prolonged exposure of a TiN film to the Al flux does not result in the  incorporation of Al, as verified \emph{ex situ} by X-ray photoelectron spectroscopy of the TiN surface. It is therefore unlikely that Al plays a role in the formation of the \{111\} TiN facets during the AlN incubation time.
	
	The morphology of the AlN NWs overgrown on the TiN films is found to depend on the annealing conditions chosen for the TiN. As displayed in Fig.\,\ref{fig1}(d), thick NWs are obtained when growth proceeds on the TiN film with large grains. Thin hexagonal NWs are only observed on the TiN film annealed under N. We conclude that the AlN nuclei grow radially on the \{111\} TiN facets until they reach the edges of the grains. This view is confirmed by the TEM cross-section shown in Fig.\,\ref{fig2}(a) of a thick AlN NW grown on a TiN film annealed in vacuum. It is clear that the radial growth of the AlN NW stopped when the NW base reached the TiN grain boundaries, as schematized in Fig.\,\ref{fig2}(b). 
	
	Previous measurements have revealed a metal-polarity for the AlN NWs grown on TiN \cite{Azadmand2020,Jaloustre2021}. In the case of GaN, the metal polarity does not allow NW self-assembly \cite{fernandez-garrido_2012}. Ga-polar GaN NWs are only obtained in the presence of a mask that limits the radial growth of the initial GaN nucleus. By analogy with the GaN case, we can then conclude that grain boundaries in the TiN layer are the key element inducing the formation of Al-polar AlN NWs on TiN. This means that the cross-sectional shape and diameter of the NWs are initially dictated by the shape and distribution of the TiN grains. However, a precise control of the morphology of these grains will be difficult to achieve using sputtering techniques. Perhaps, a more favorable option would be to grow the AlN NWs with N polarity, to promote the appearance of NW sidewalls without the need for \textit{in situ} masking. N-polar AlN could be obtained by changing the surface termination of the TiN film, \textit{e.g.}\ by oxidizing its surface \cite{Mohn2016}. 
	In the following, grain growth will be mitigated by systematically annealing the TiN under N exposure in order to obtain the thinnest possible AlN NWs.
	
	\subsection{Kinetics of AlN NW growth}\label{subsec:AlN-kinetics}
	During MBE growth of binary nitrides, a large adatom diffusion length for the group III atoms is generally desired to improve the structural quality of the growing crystal \cite{neugebauer_2003}. This is even more important for NW self-assembly where adatom diffusion on the NW sidewalls substantially contributes to the NW axial elongation rate \cite{Debnath2007,songmuang_2010,Galopin2011,Consonni2012}. Besides increasing adatom diffusion lengths, a high growth temperature is beneficial to decrease the density of self-assembled NWs \cite{zettler_2015}. 
	
	\begin{figure*}
		\includegraphics[width=\textwidth]{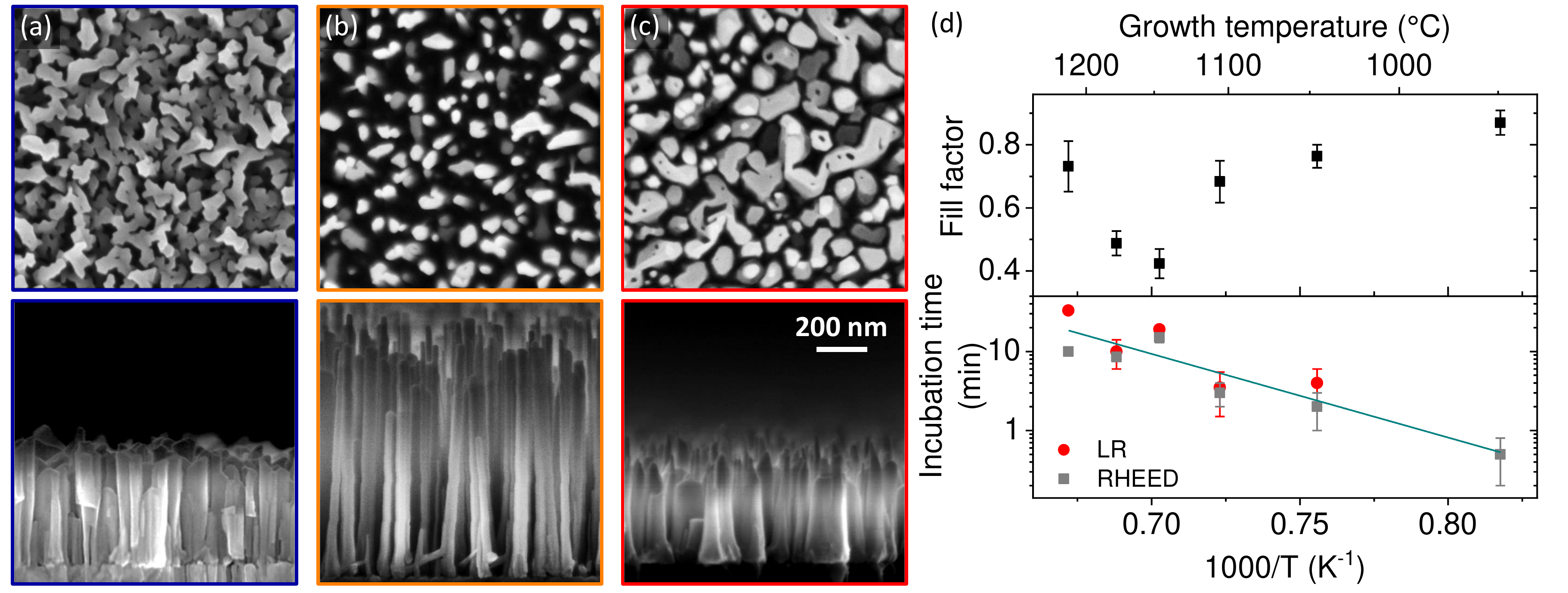}
		\caption{Top-view (first row) and cross-section (second row) SEM views of AlN NW ensembles grown on TiN annealed under N exposure at (a) 950\degC{}, (b) 1150\degC{}, and (c) 1215\degC{}. The scale is identical for all images and indicated in (c). The NW incubation times and fill factors as a function of growth temperature are displayed in (d).}\label{fig3}
	\end{figure*}
	
	In the following, we grow AlN NWs in a temperature range from 950 to 1215\degC{} on TiN films annealed under N exposure. Top-view and cross-section SEM images of the obtained NW ensembles are shown in Fig.\,\ref{fig3}. 
	Surprisingly, changing the growth temperature by 250\degC{} does not induce a drastic change in the NW density. By increasing the substrate temperature from 950 to 1150\degC{}, the NW fill factor is reduced by at most a factor of 2 and the incubation time increases moderately from 0 to 15\,min, with an apparent activation energy of 2.1\,eV [Fig.\,\ref{fig3}(d)]. In comparison, the growth window of self-assembled GaN NWs on Si is narrower ($\approx 50$\degC{} \cite{fernandez-garrido_2009}) and heavily coalesced ensembles are typically obtained for such short incubation times observed here. As already discussed in our previous work \cite{Azadmand2020}, this finding likely results from the reduced propensity of AlN NWs to coalesce by bundling \cite{Kaganer2016}.
	
	At 950\degC{} [Fig.\,\ref{fig3}(a)], the NW ensembles are similar in morphology to those grown on SiN$_x$ and SiO$_x$ films \cite{Landre2010,E2015,Gacevic2021}. The observed inverse tapering of the NWs is also characteristic for the lack of diffusion between the NW sidewalls and top facet \cite{Daudin2020}. By increasing the growth temperature, NW coalescence reduces along with the decrease in fill factor [Fig.\,\ref{fig3}(b)]. Yet, above 1200\degC{}, the pronounced grain growth occurring in the underlying TiN layer prior to AlN nucleation results in wider NWs, that are similar in shape as the coalesced ones grown on vacuum annealed TiN [Fig.\,\ref{fig3}(c)], eventually leading to an increase in fill factor despite an increase of incubation time to around 20\,min. A further increase in substrate temperature would probably help lengthen the incubation time, with consequences on the NW density \cite{zettler_2015,Kaganer2016}, but the concomitant formation of large TiN \{111\} facets will eventually result in the formation of AlN blocks instead of thin NWs. The TiN grain growth thus seems to hinder the possibility of achieving ultra-low densities of AlN NWs as obtained for GaN on TiN \cite{Auzelle2022}.
	
	\begin{figure*}
		\includegraphics[width=\textwidth]{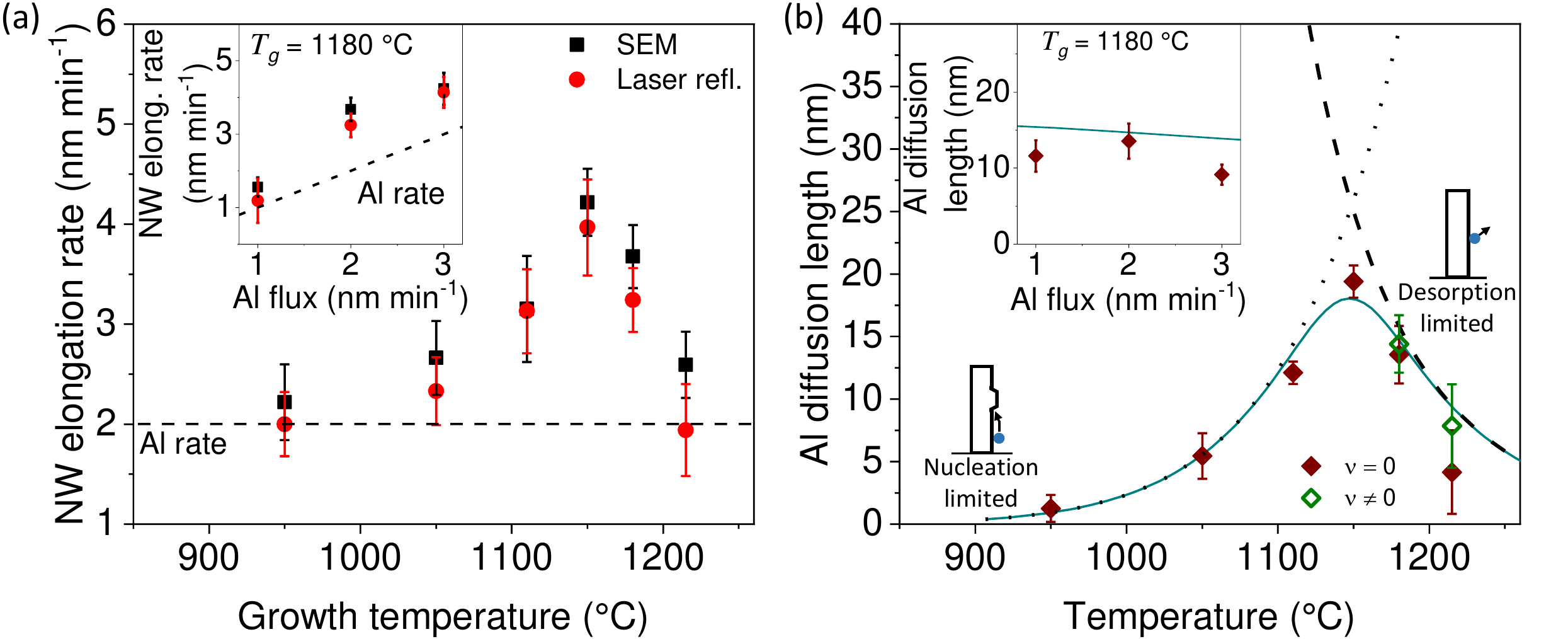}
		\caption{(a) NW mean elongation rate as a function of growth temperature determined by cross-section SEM and laser reflectance. The inset shows the NW elongation rate as a function of Al flux. (b) Al diffusion length $\lambda$ on the NW sidewalls (along the [0001] direction) as a function of temperature, extracted from Eq.\,(\ref{eqn:rate}) with $g = 0.8$. In the case of  $\nu \neq 0$, the AlN decomposition kinetics from Ref. \cite{fan_2001} are used in Eq.\,(\ref{eqn:rate}), i.e. $\nu = C \exp(-E_{dec}/k_{B}T)$ with $C =1.4 \times 10^{18}$ and $E_{dec} =5.4$\,eV. The dotted and dashed curves illustrate the two different mechanisms limiting the Al diffusion length. Simulations according to Eq.\,(\ref{eqn:desorb2}) with $E_\mathrm{des} = 6.0$\,eV are shown by solid lines.}
		\label{fig4}
	\end{figure*}
	By varying the substrate temperature and the Al flux, NWs of very different lengths are obtained for similar growth times. The mean NW elongation rate extracted from cross-sectional SEM images and from the oscillation period in the laser reflectance signal is plotted as a function of growth temperature and Al flux in Fig.\,\ref{fig4}(a). It evidences a non-monotonic dependence of the growth rate on the substrate temperature, similar to the GaN NW case \cite{Consonni2012}. This behaviour is associated to variations in the Al adatom diffusion length $\lambda$ at the NW sidewalls along the [0001] direction, so that the NW elongation rate $dL/dt$ can be described as follows \cite{Consonni2012}:
	
	\begin{equation}
		\frac{dL}{dt} = \Phi_\mathrm{Al} \left (1 -\nu + \frac{4\lambda}{\sqrt{3}R}g\tan{\alpha} \right ) \label{eqn:rate}
	\end{equation}
	with $\Phi_\mathrm{Al}$ the flux of Al atoms impinging on the NW top facet, $\nu$ the desorbing  flux of Al atoms from the top facet, $R$ the NW radius, $g \leq 1$ reflecting the difference in chemical potential of Al between the top facet and the sidewalls \cite{dubrovskii_2009}, and $\alpha$ the angle between the Al flux and the NW axis (here, $\alpha = 38$\,$^\circ{}$). 
	This model applies to Al-limited growth and accounts for deposition/desorption of Al on the NW top facet as well as diffusion of Al adatoms from the NW sidewalls to the top facet. 
	
	Due to the strongly N-rich conditions, we expect immediate binding of Al impinging on the top facet with N adatoms and therefore assume $\nu=0$ for simplicity. 
	In contrast to the NW top facet, the sidewalls are only exposed alternately to the Al and N fluxes, respectively, providing a chance for Al to diffuse. Taking $g = 0.8$ (as for GaN NWs \cite{Consonni2012}) and the average NW radius of each ensemble determined from top-view and cross-section SEM images, values of $\lambda$ are extracted from Eq.\,\ref{eqn:rate} and plotted as a function of $T$ and $\Phi_{Al}$ in Fig.\,\ref{fig4}(b). Similar to the growth rate, $\lambda$ increases with temperature up to $1150$\degC{} and decreases above. As detailed in Ref.\,\cite{Consonni2012}, this non-monotonous trend can be described using two exponential terms accounting for desorption and incorporation of Al adatoms at the NW sidewalls, respectively:
	

	%
	
	\begin{equation} \label{eqn:desorb2}
		\lambda = \left [  A \exp{ \left ( -\frac{ E_\mathrm{des} - E_\mathrm{diff} }{k_\mathrm{B}T} \right ) } + 
		B \Phi_\mathrm{Al}^{3/2}\exp{ \left ( \frac{E_\mathrm{nucl}}{k_\mathrm{B}T} \right )} \right ] ^{-\frac{1}{2}}\\	 
	\end{equation}
	where $E_\mathrm{des}$ is the energy barrier for Al desorption from the NW sidewall, $k_\mathrm{B}$ the Boltzmann constant, $E_\mathrm{diff} = 0.6$\,eV \cite{Jindal2009} the diffusion barrier for Al adatoms on the NW sidewalls obtained by \emph{ab initio} calculations, $E_\mathrm{nucl}$ the barrier for AlN nucleation at the NW sidewalls, and $A$ and $B$ are fitting parameters. 
	
	The low-temperature side of the experimental data is well reproduced by taking $E_\mathrm{nucl} = 5.0$\,eV, similar to the GaN case \cite{Consonni2012}. In this temperature range, $\lambda$ is then limited by the competing radial growth of the NWs. Above $1150$\degC{}, $\lambda$ is limited by thermal desorption of Al adatoms from the NW sidewalls. However, no satisfactory fit can be obtained using $E_\mathrm{des}<(10\pm1)$\,eV, which is much higher than the decomposition energy of bulk AlN ($\approx 6$\,eV   \cite{fan_2001,Dreger1962}). This suggests that the initial assumption of $\nu = 0$ is invalid in the high-temperature range. For more accurate modeling, we use in the following $\nu = C \exp(-E_{dec}/k_{B}T)$ with $C =1.4 \times 10^{18}$ and $E_{dec} =5.4$\,eV, as shown in Fig.\,S2 of the SI, which corresponds to the decomposition rate of AlN measured in vacuum \cite{fan_2001}. The updated values of $\lambda$ are shown by the open symbols in Fig.\,\ref{fig4}(b). Compared to the initial case, $\nu =0$, a significant change in diffusion length is only observed above 1150\degC{} and $\lambda$ can now be modeled with an Al desorption energy $E_\mathrm{des} = (6.0\pm 1.0$)\,eV. This value is still much larger than the desorption energy of Al from AlN(0001) surfaces (3.8\,eV \cite{Lee2020}), suggesting that no Al adlayer forms at high temperature at the NW sidewalls. Instead, Al would reach the top facet by successive adsorption and desorption events between neighboring NWs, as already observed for GaN NW growth \cite{sabelfeld_2013}. In this process, binding between Al and N adatoms at the NW sidewalls would be the mechanism limiting the Al mass transport to the top facet, in agreement with an activation energy equal to the decomposition energy of AlN (5.4 -- 6.3\,eV \cite{fan_2001,Dreger1962}). As a result, the mass transport at high temperature is intimately related to the sample morphology, substrate rotation speed and cell arrangement, which would all need to be taken into account for accurate modeling of the high-temperature AlN NW growth. 
	
	
	Here, (1150$\pm$20)\degC{} appears as the optimal growth temperature to favor NW elongation, which is about 100\degC{} larger than for homoepitaxy of AlN layers on bulk substrates \cite{Lee2020,Singhal2022}. Strikingly, $\lambda$ is negligibly small below 950\degC{}, the temperature range in which AlN NWs on a GaN NW template are fabricated. Hence, for these NWs, the ratio between vertical and radial growth is solely governed by geometrical parameters of the MBE chamber \cite{foxon_2009}, which typically leads to tapered NWs \cite{Galopin2011}. Such NWs should also exhibit large densities of native point defects.

	\subsection{Point defect incorporation}\label{subsec:CL}
	
	An increase in the adatom diffusion length is typically expected to reduce point defect incorporation. In the following, we examine the band-edge and deep-level CL of the self-assembled AlN NWs to investigate point defect incorporation. 
	
\begin{figure*}
		\includegraphics[width=\textwidth]{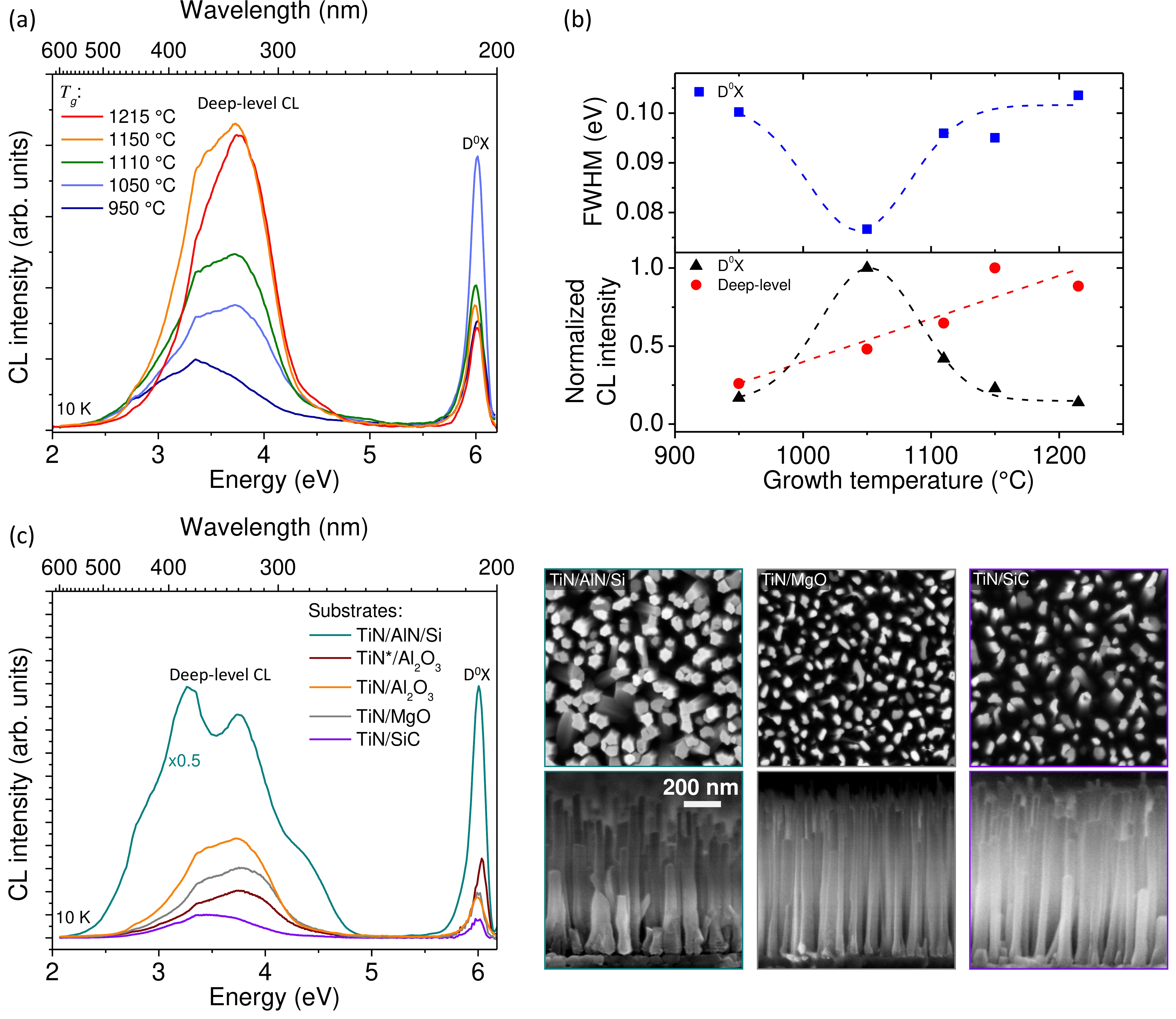}
		\caption{(a) CL spectra of AlN NWs grown on TiN/Al$_2$O$_3$ between 950 and $1215$\degC. (b) Normalized intensities of the $D^{0}X$ and the deep-level luminescence, and FWHM of the $D^{0}X$ luminescence. (c) CL spectra of AlN NWs grown on various substrates at $\approx$1150\degC{}. TiN annealed in vacuum is marked by an asterisk. The morphologies of NWs grown on the different substrates are shown in the top-view and cross-section SEM images next to (c).}
		\label{fig5}
	\end{figure*}

	CL spectra acquired at 10\,K on the AlN NW ensembles grown on TiN/Al$_2$O$_3$ between 950 and 1215\degC{} are displayed in Fig.\,\ref{fig5}(a). They feature a band-edge recombination at 5.99--6.02 eV associated to donor bound excitons ($D^{0}X$) and a deep-level luminescence consisting of two contributions at around 3.4 and 3.8 eV. Deep-level luminescence in AlN is commonly observed in the presence of Al vacancies and oxygen impurities \cite{Koppe2016}. We note that the N-rich conditions used here likely favor the formation of Al vacancies \cite{Yan2014,Gaddy2013,Hevia2013,Stampfl2002} and that O is a common background impurity in self-assembled GaN NWs grown by MBE \cite{zettler_2015a}. No intense transitions are seen around 5\,eV as sometimes reported for AlN NWs grown on GaN stems \cite{wu_2020,vermeersch_2023}. The band-edge is 75--105\,meV wide, which is narrower than for AlN NWs grown on GaN stems \cite{wu_2020,vermeersch_2023} but broader than for direct MBE growth of AlN NWs on Si substrates ($20 $\,--\,$ 50$\,meV) \cite{Landre2010,Gacevic2021}, and than for vapor phase epitaxy of strain-free AlN layers ($\approx 1$\,meV) \cite{Gacevic2021}). We associate the broad transition in our AlN NWs to the presence of a large density of ionized donors \cite{schubert_1997}. 
	
	Remarkably, the intensity of the deep-level luminescence is seen to increase as a function growth temperature, with the contribution at 3.8 eV becoming more dominant. In contrast, the band-edge luminescence reaches a maximum for growth at 1050\degC{} and then decreases with temperature, as depicted in Fig.\,\ref{fig5}(b). The band-edge transition also features its smallest width at 1050\degC{}. Incorporation of ionized impurities thus seems to reach a minimum at 1050\degC{} and further increases above. An increased incorporation of Al vacancies directly above 1050\degC{} is unlikely since Al desorption was seen to become dominant only above 1180\degC{}. Major intermixing between TiN and AlN is not visible in the TEM images of Fig.\,\ref{fig2}(a). Thus, we propose that the increased deep-level luminescence above 1050\degC{} results from O incorporation from the Al$_2$O$_3$ substrate. Interfacial solid-state reactions have been previously seen for sputtered TiN deposited on MgO substrates already at 850\degC{} \cite{hultman_2000} and a similar reaction can be expected with sapphire. This is substantiated by the presence of voids at the Al$_2$O$_3$/TiN interface after growth at $1215$\degC{}, as shown in Fig.\,S2 of the SI. O diffusion from Al$_2$O$_3$ into the grain boundaries of sputtered AlN layers was also reported, although at higher temperature ($>1500$\degC{}) \cite{Cancellara2021}. Here, the onset of O diffusion ($>1000$\degC{}) would correlate with the beginning of the TiN layer transformation. The O released from the Al$_2$O$_3$ substrate will diffuse along the TiN grain boundaries before incorporating into the AlN NWs. A higher O content is expected at high growth temperature, which will result in the formation of DX centers instead of neutral shallow donors \cite{gordon_2014}. The presence of such DX centers can thus explain the decrease in the band-edge luminescence intensity above $1050$\degC{}, since excitons do not bind to DX centers \cite{Koppe2016}.
	
	To test for this hypothesis, additional AlN NWs are grown at $\approx$1150\degC{} on TiN films sputtered on Si(111), MgO(111), and SiC(0001) substrates. Solid-state reactions at the SiC/TiN interface should be negligible up to 1100\degC{} \cite{hultman_2000}. In contrast, strong interdiffusion is observed at the Si/TiN interface, with deleterious consequences on the AlN NW morphology (Fig.\,S3 in the SI). To mitigate this effect, a 20\,nm thick AlN layer is grown by MBE on the Si substrate prior deposition of a $65$\,nm thick TiN film. Growth on this substrate results in vertical AlN NWs, but the presence of voids in the silicon substrate indicates that Si diffusion in the TiN layer could not be entirely avoided. As a matter of fact, the NWs exhibit a star-like cross-section, similar to heavily Si-doped GaN and AlN NWs \cite{Fang2015,Vermeersch2021}. 
	
	The CL spectra of AlN NWs grown on the different substrates are shown in Fig.\,\ref{fig5}(c). The SEM micrographs next to the spectra confirm that the NW ensembles grown on MgO and SiC have similar morphologies to those grown on Al$_2$O$_3$ [cf. Fig.~\ref{fig3}(b)]. All samples exhibit a similar ratio between band-edge and deep-level luminescence but drastically different overall CL intensities. The most intense signal is obtained for NWs grown on the Si substrate. Due to the outdiffusion of Si, these NWs are likely to be heavily doped and may even exhibit a SiN$_x$ shell \cite{Fang2015}. AlN NWs grown on Al$_2$O$_3$ and MgO substrates exhibit essentially similar CL intensities. On the contrary, the weakest signal is observed for NWs grown on SiC, where outdiffusion is negligible. Therefore, heavy Si or O doping appears to be beneficial for enhancing the NW CL intensity, which is counterintuitive since these donors form DX centers at high concentration \cite{gordon_2014,vermeersch_2022}. We thus propose that the enhancement of the CL intensity results from a reduction of the nonradiative recombination at the NW surface, either due to a passivation of the surface states by the partial formation of SiN$_x$ and AlO$_x$ shells \cite{lu_2015,wang_2017}, or by a screening of the surface electric field responsible for the field-ionization of the excitons \cite{wang_2017,auzelle_2021}. Along this line, thicker NWs should suffer less from nonradiative surface recombination, in agreement with the observed higher band-edge CL of AlN NWs grown on vacuum annealed TiN/Al$_2$O$_3$  [marked by an asterisk in Fig.\,\ref{fig5}(c)] compared to those grown on TiN/Al$_2$O$_3$ annealed under N exposure. 
	
	A high incorporation of impurities such as O can also introduce additional defects like inversion domains \cite{Stolyarchuk2018,Bruley1994,Gorzawski1995,Youngman1989,Westwood1991,Stolyarchuk2017,Mohn2016}. Such defects are indeed observed by TEM in many AlN NWs grown on the vacuum annealed TiN film. As shown in Fig.\,\ref{fig2}(c), the very thin inversion domains are identified in the dark field mode, based on the contrast reversal occurring in different Bragg imaging conditions \cite{Serneels1973,Romano1996}. High-resolution images [Fig.\,\ref{fig2}(d)] also reveal the expected vertical shift by half a unit cell compared to the surrounding matrix. Similar defects have been observed for NWs grown on TiN annealed under N (data not shown). Most of the inversion domains have diameters between 2 to 10\,nm, propagate nearly vertically, and eventually close, as shown in Fig.\,S4 of the SI. O is a likely source for their nucleation, but Ti cannot be ruled out either \cite{kong_2016}.

	\section{Conclusions}\label{sec:Con}
	
	We have examined the kinetics of AlN NW growth on sputtered TiN between $950$ and $1215$\degC{} and the TiN stability in this temperature range. Above $1100$\degC{}, the sputtered TiN films undergo a reshaping process involving grain growth and the formation of \{111\} facets, which is presumably enabled by the accumulation of N vacancies at the TiN surface. AlN nucleates preferentially on the \{111\} facets and laterally extends until reaching the grain boundaries. It follows that the AlN NW cross-section is initially determined by the size of the underlying TiN grains. Modeling of the mean NW elongation rate shows that the Al adatom diffusion length maximizes at ($1150\pm20$)\degC{}, reaching a value of around 20\,nm. Regardless of the used substrate, $1150$\degC{} thus appear as the optimal growth temperature for MBE of AlN NWs. 
	However, at this temperature, the fabricated AlN NWs exhibit an intense deep-level CL and a broad donor-bound exciton transition, which we associate to the presence of a large density of ionized impurities. Increasing the substrate temperature from $950$ to $1215$\degC{} actually increases the deep-level luminescence, an effect that is attributed to the incorporation of O released from the Al$_2$O$_3$ substrate and diffusing through the TiN grain boundaries. Comparison of the luminescence intensity of AlN NWs grown on Si, MgO and SiC substrates suggests that heavy Si and O doping of the AlN NWs due to interdiffusion at the TiN/substrate interface increases the NW internal quantum efficiency. This effect may originate from the formation of a SiN$_x$ or an AlO$_x$ shell acting as a passivating layer. The observation of inversion domains in the AlN NWs is also attributed to O and Si impurities which have segregated at the TiN surface.
	
	For the fabrication of AlN NW ensembles suitable for optoelectronic and piezoelectric applications, substrate degradation above 1100\degC{} must be avoided. This can be achieved by growing on SiC and Al$_2$O$_3$ substrates buffered with AlN or, possibly, by replacing the TiN nucleation layer by more stable transition metal nitrides like ZrN, HfN or NbN \cite{lengauer_2015}. Moreover, a surface passivation of the AlN NWs such as given by SiN$_x$ or AlO$_x$ coatings appears necessary to enhance their radiative efficiency.

	\section{Acknowledgement}
	
	The authors thank Carsten Stemmler for maintenance of the MBE system, Anne-Kathrin Bluhm for SEM imaging and Doreen Steffen for TEM sample preparation. The authors are grateful to Oliver Bierwagen for providing MgO substrates. We thank Georg Hoffmann for a critical reading of the manuscript and Markus Wagner for fruitful discussions. Funding from Deutsche Forschungsgemeinschaft and Agence Nationale de la Recherche through the project Nanoflex (ANR-21-CE09-0044) is gratefully acknowledged.

	\section{Supporting Information}

	The Supporting Information is available free of charge and contains the following sections:
	Laser reflectance of growing AlN NWs and extraction of the NW mean elongation rate; decomposition kinetics of AlN; cross-section SEM images of AlN grown on TiN/Si, TiN/AlN/Si and TiN/Al$_2$O$_3$ showing interdiffusion; TEM micrograph of an AlN NW showing IDs.

	\bibliographystyle{iopart-num}
	
	\bibliography{References}
	
\end{document}


\title[Supporting Information: High temperature MBE of AlN NWs]{Supporting Information: Growth kinetics and substrate stability during high-temperature molecular beam epitaxy of AlN nanowires}
	
	\author{P.\,John$^1$, M.\,G\'{o}mez Ruiz$^1$, L.\,van Deurzen$^2$,  J.\,L\"ahnemann$^1$, A.\,Trampert$^1$, L.\,Geelhaar$^1$, O.\,Brandt$^1$, and T.\,Auzelle$^1$ }
	
	\address{$^1$ Paul-Drude-Institut f\"ur Festk\"orperelektronik, Leibniz-Institut im Forschungsverbund Berlin e.V.,	Hausvogteiplatz 5–7, 10117 Berlin, Germany}
	\address{$^2$ School of Applied and Engineering Physics, Cornell University, 14853 Ithaca New York, USA}
	
	\ead{john@pdi-berlin.de}

	%

	%
	\ioptwocol

	
	\section{Growth kinetics}
	
	In the following is discussed, how the incubation time and the nanowire (NW) elongation rate are obtained by coupling \emph{in situ} reflective high energy electron diffraction (RHEED) and laser reflectance measurements with \emph{ex situ} scanning electron microscopy measurements (SEM). Further, we explain our estimation of $\nu$ in Eq.\,(1), the rate of Al loss from the NW top facet for the diffusion model of the main manuscript. 
	
	\subsection{Incubation time}
	\label{subsec:inc-time}
	
\begin{figure}[h]
		\includegraphics[width=\columnwidth]{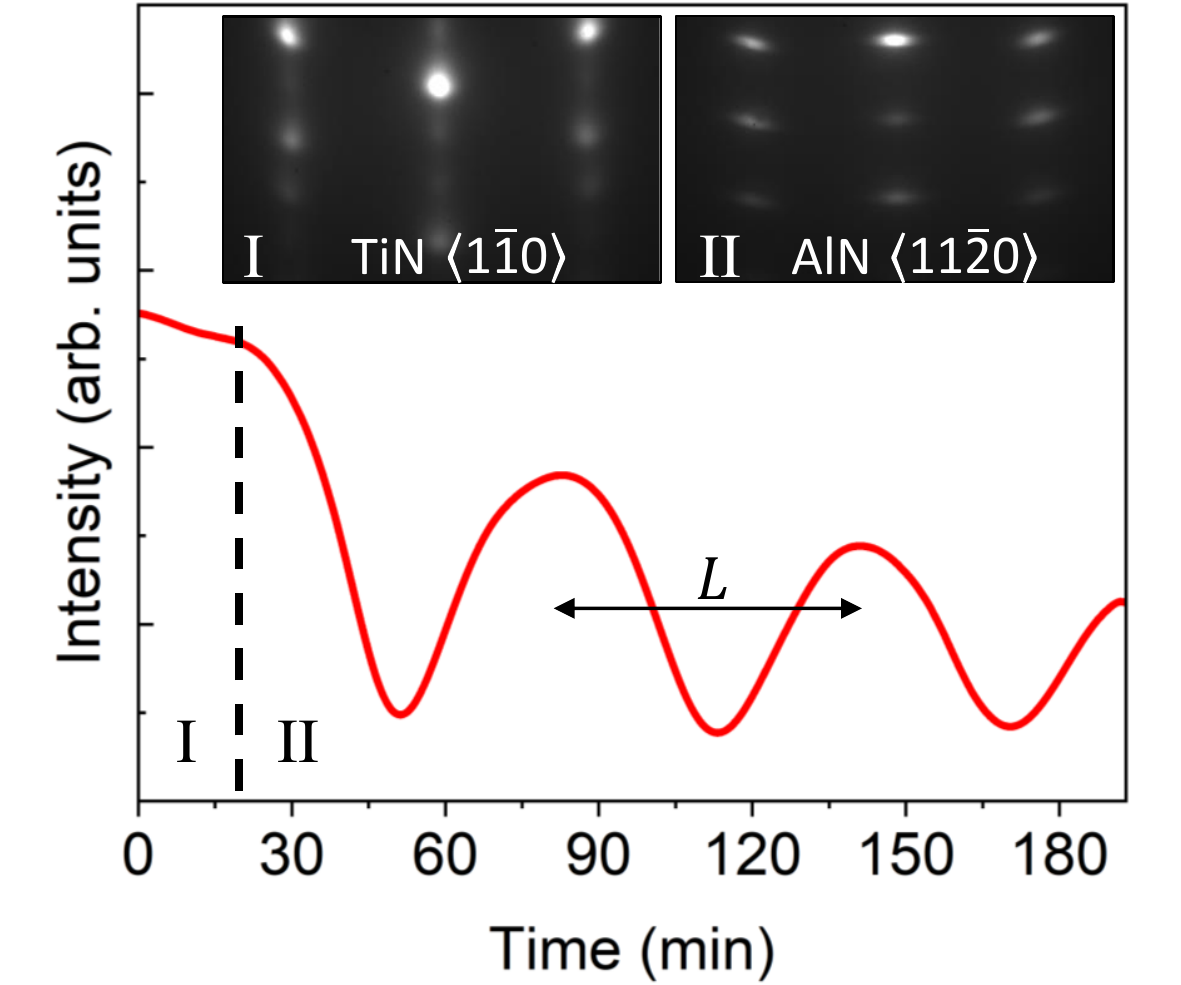}
		\caption{Laser reflectance signal as a function of time. The Al and N shutters are opened at $0$\,min, the dashed line indicates the onset of NW nucleation. Insets (I) and (II) show the RHEED pattern of a nitridized TiN surface and an AlN ensemble, revealing (111) and (0001) orientations \cite{Wolz2015}, respectively.}
		\label{fig:LR}
	\end{figure}

	The incubation time $t_{inc}$ is determined for each NW ensembles by RHEED and laser reflectance. In the first case, the transition between the TiN(111) to the AlN(0001) RHEED patterns that are exemplified in the insets (I) and (II) of Fig. \ref{fig:LR}(a) is used to monitor the onset of AlN nucleation. In the second case, the beginning of signal oscillation in laser reflectance is used to determine the incubation time [see dashed line in Fig. \ref{fig:LR}(a)]. The values from both methods are compared in Fig.~3(d) of the main manuscript and follow an Arrhenius behavior, similar to the self-assembled GaN NW case \cite{Consonni2011a}.

	\subsection{NW elongation rate}
	The NW elongation rate is first obtained by dividing the average NW length extracted from cross-section SEM measurements by the NW growth duration ($t = t_{total} - t_{inc}$). This approach may however underestimate the growth rate since many NWs are still nucleating after the incubation time.
	
	As an alternative, the NW growth rate can be determined from the period of the oscillations in the laser reflectance signal. During a period $L$, the effective layer thickness has increased by an amount $w$:
	
	\begin{equation}
		w = \frac{\lambda}{2 n_{eff} \cos \beta}, \label{eqn:length}
	\end{equation}
	where $\lambda=650$\,nm is the laser wavelength, and $\beta$ is the angle of light propagation in the growing material, linked to the laser incident angle $\alpha=74^{\circ{}}$ by Snell's law. 
	In the case of AlN NW ensembles, we use an effective refractive index $n_{eff} = 1 + f(n_{AlN} - 1)$ \cite{Janicki2008,Liu2017}, with $n_{AlN} = 2.1$ and $f$ the NW fill factor determined by top-view SEM measurements. The NW growth rate eventually amounts to $dL/dt = w/L$.
	
	\subsection{AlN decomposition}
\begin{figure}[h]
		\includegraphics[width=\columnwidth]{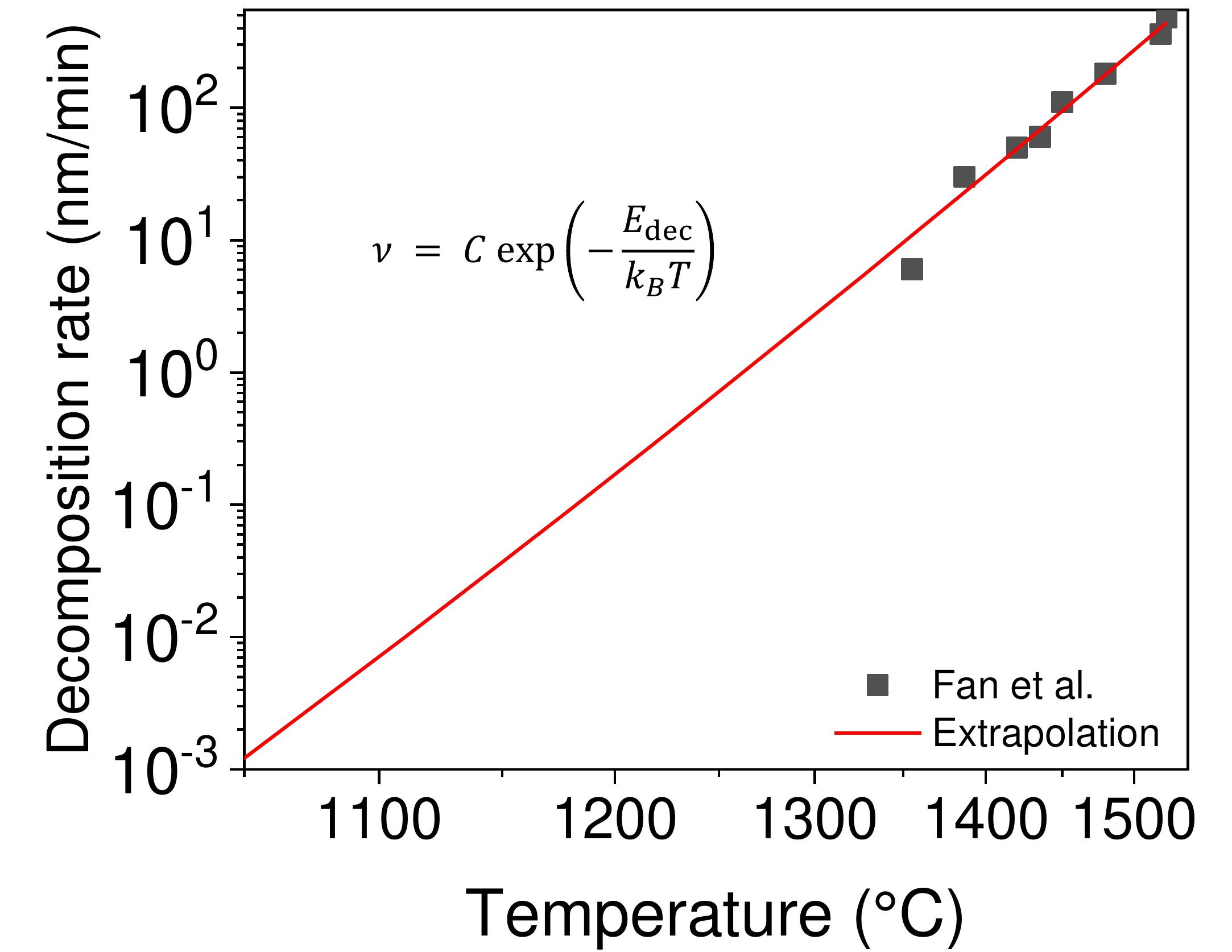}
		\caption{AlN decomposition rate as found by Fan et al., extrapolated to lower temperatures. }
		\label{fig:dec}
	\end{figure}

	The diffusion model discussed in Section\,3.2 of the main manuscript contains a term describing the loss of Al adatoms from the NW top facet. Due to strongly N-rich conditions, the loss of Al by desorption requires to break AlN bonds with an activation energy close to the decomposition energy of AlN. We hence extrapolate the decomposition rate of bulk AlN found by Fan et al.~in the range of 1350 to 1530\degC{} to the growth temperature range used for our experiments and use this data to estimate the loss of Al $\nu$ from the NW top facet in Eq.\,(1) of the main manuscript. As shown in Fig.\,\ref{fig:dec}, the decomposition rate follows a temperature activated behavior, where $\nu = C \exp(-E_{dec}/k_{B}T)$ with $C =1.4 \times 10^{18}$ and $E_{dec} =5.4$\,eV. Significant decomposition is hence only observed for the AlN NWs grown at 1180 and 1215\degC{} with rates of 0.26 and
	0.09\,nm/min, respectively. 
	
	\section{Interfacial reactions with TiN}
	
\begin{figure}[h]
		\includegraphics[width=1\columnwidth]{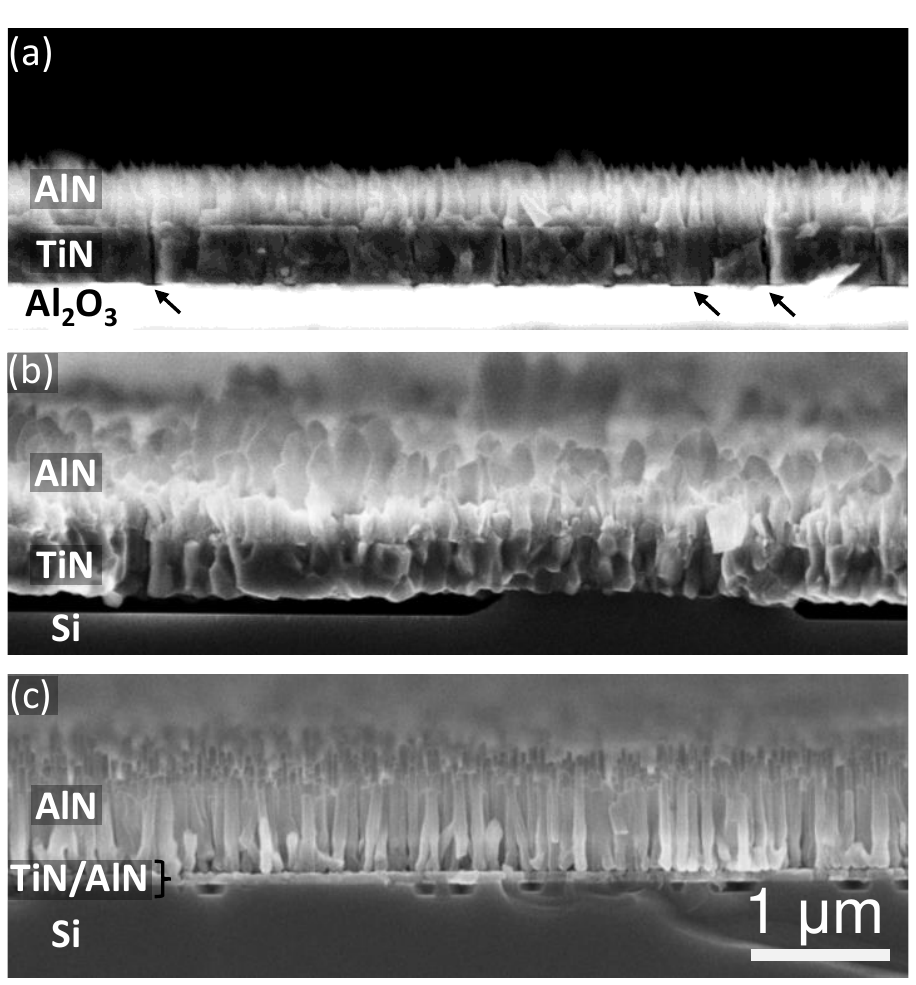}
		\caption{Cross-section SEM images of AlN grown on (a) TiN/Al$_2$O$_3$ at 1215\degC{}, (b) TiN/Si and (c) TiN/AlN/Si at 1150\degC{}. Voids at the interface between the substrate and the TiN layer evident interdiffusion of substrate atoms that are eventually incorporated in the NWs.}
		\label{fig:SEM}
	\end{figure}

	The interfacial reactions between TiN and the underlying substrate become visible in cross-section SEM. As shown in Fig.\,\ref{fig:SEM}(a), small voids are formed at the TiN/Al$_{2}$O$_{3}$ interface. These voids are preferentially formed at TiN grain boundaries (see black arrows), pointing out their role as diffusion channel. The released oxygen atoms may eventually be incorporated in the growing AlN NWs.
	The interdiffusion becomes even more severe at the TiN/Si interface. Large voids are found, indicating a massive outdiffusion of Si, which leads to a degradation of NW morphology [Fig.\,\ref{fig:SEM}(b)]. By growing a thin AlN buffer layer prior TiN sputtering, interfacial reactions are reduced. NWs with a star-like cross-section are grown, indicating heavy Si doping or the formation of a thin SiN$_{x}$ passivation shell.

	\section{IDs in AlN NWs}
\begin{figure}[h]
		\includegraphics[width=1\columnwidth]{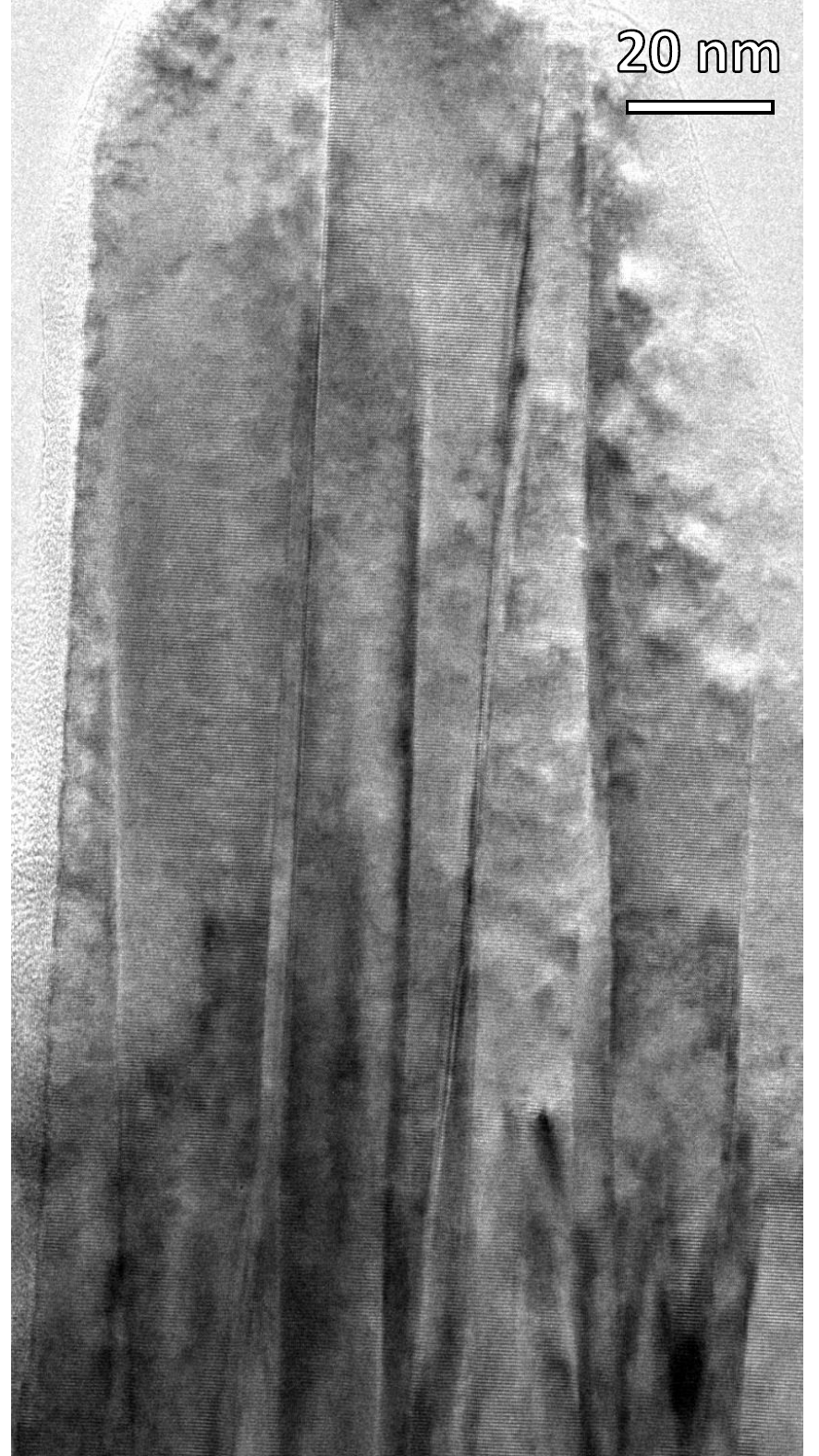}
		\caption{Cross-section TEM image of an AlN NW with numerous IDs, penetrating the NW.}
		\label{fig:TEM}
	\end{figure}

	The outdiffusion of O from the Al$_{2}$O$_{3}$ substrate is a likely source for IDs in our AlN NWs, as explained in Fig.\,2 of the main manuscript.
	To get a better insight into the distribution and number densities of IDs in AlN NWs, an additional TEM image of a NW grown on annealed TiN/Al$_{2}$O$_{3}$ is shown here. The image in Fig.\,\ref{fig:TEM} reveals that numerous IDs are formed in one NW. These originate at the interface to the TiN surface and propagate through the NW. While for thin NWs the defective region is confined to the TiN interface, as reported in Ref. \citenum{azadmand2020}, only some of the IDs close during growth in the case of thicker NWs.
	
	\bibliographystyle{iopart-num}
	
	\bibliography{AlNNWs_PJ}